\documentclass[12pt]{article}

\usepackage[titletoc,toc,title]{appendix}
\usepackage{graphicx}
\usepackage{dcolumn}
\usepackage{bm}
\usepackage{subfig}
\usepackage{caption}
\usepackage{epsfig}
\usepackage{epstopdf}
\usepackage{natbib}
\usepackage{graphics}
\usepackage{latexsym}
\usepackage{amsmath}
\usepackage{calrsfs} 
\usepackage{rotating}
\usepackage{authblk}
\usepackage{xfrac}
\usepackage{color}
\usepackage{ifpdf}

\newcommand{\ie}{{\it i.e.}}
\newcommand{\eg}{{\it e.g.}}

\graphicspath{{./pics/}}
\date{}

\begin{document}

\title{Drop deposition on surfaces with contact-angle hysteresis: Liquid-bridge stability and breakup}
\author{Amir Akbari and Reghan J. Hill\footnote{E-mail: reghan.hill@mcgill.ca}}
\affil{Department of Chemical Engineering, McGill University, Montreal, Quebec H3A 0C5}

\maketitle

\begin{abstract}
We study the stability and breakup of liquid bridges with a free contact line on a surface with contact-angle hysteresis under zero-gravity conditions. Theoretical predictions of the stability limits are validated by experimental measurements. Experiments are conducted in a water-methanol-silicon oil system where the gravity force is offset by buoyancy. We highlight cases where stability is lost during the transition from a pinned-pinned to pinned-free interface when the receding contact angle is approached---rather than a critical state, indicating that the breakup length is not always associated with the static maximum-length stability limit. We demonstrate that the dynamic contact angle controls the contact-line radius following stability loss, and that interface evolution following stability loss can increase the dispensed-drop size if the contact angle is fixed.
\end{abstract}

\section{Introduction} \label{sec:introduction}

Studying the stability and dynamics of liquid bridges is motivated by a broad range of applications, including crystal growth in microgravity~\citep{velarde1988physicochemical}, surface patterning, nano-printing, and nano-lithography~\citep{salaita2007applications, huo2008polymer, shim2011hard}, aggregation and coalescence of flexible fibres~\citep{cohen2003kinks, kim2006capillary, boudaoud2007elastocapillary, pokroy2009self}, and capillary induced collapse of elastic structures~\citep{mastrangelo1993mechanical, kwon2008equilibrium, chandra2009capillary, chini2010understanding, akbari2014an, akbari2014stability}. Quantifying liquid-bridge and jet breakup upon stability loss dates to the works of \cite{plateau1873statique} and \cite{rayleigh1879capillary, rayleigh1879instability}. While early investigations on crystal growth and purification focused on determining the minimum liquid volume that can be held between circular discs~\citep{martinez1986liquid, meseguer1995review}, the drop-size distribution following breakup is of prime interest in contact-drop dispensing and liquid-transfer applications~\citep{dodds2009stretching, lutfurakhmanov2010capillary, chen2014liquid}. Minimizing the dispensed-drop size relative to the needle diameter is central to surface patterning based on direct-write lithographic techniques~\citep{piner1999dip, huo2008polymer}.

Recent studies on contact-drop dispensing have shown that the deposited drop size is influenced by the needle retraction speed, needle-tip size, surface characteristics, and dispensing control parameters~\citep{qian2009micron, qian2011motion}. Interestingly, the deposited drop volume in pressure-controlled and volume-controlled dispensing behave differently with the needle retraction speed. Faster retraction reduces the drop size to a minimum and monotonically increases the drop size in pressure-controlled and volume-controlled dispensing, respectively. Three regimes were experimentally identified with respect to the retraction speed $U_{n}$ for the pressure-controlled case. In the first two, $U_{n}\ll u_w$, where $u_w$ is the capillary-wave speed; the contact line is advancing in the first and stationary in the second, and the drop size scales as $U_{n}^{-1/2}$; the third corresponds to fast retraction speeds ($U_{n}/u_{w}\sim O(10^{-2}$)) where the dynamics dramatically change, and the drop size does not scale with $U_{n}$ as a simple power law. Here, the drop size is almost two orders of magnitude smaller than in the first two regimes, which \citet{qian2009micron} attributed to a fast receding contact line with a speed approaching $u_{w}$. However, in volume-controlled deposition, the dispensed-drop size did not exhibit the same sensitivity to the needle retraction speed in the parameter range studied by \citet{qian2011motion}. Thus, our study is motivated, in part, by seeking to answer whether it is possible to influence---by purely geometric means---the dynamics in volume-controlled deposition, so that one may achieve comparable sensitivity as in the pressure-controlled case, to achieve small-drop deposition.

Liquid transfer in contact-drop dispensing is generally accomplished by the breakup of a drop bridging a sharp-edged needle and a flat substrate. The bridge is pinned to the needle edge, and may have a free or pinned contact line with the substrate, depending on the surface wettability. To help distinguish the effects of geometric parameters on the dispensed-drop size from dynamics ones, \cite{akbari2014bridge} studied the static stability of liquid bridges with free-pinned contact lines, showing that the contact-line radius and bridge-neck position---key parameters determining the dispensed-drop size---can be controlled by purely geometric means. Constructing the stability region at fixed contact angles, they also showed that the free contact line has symmetry-breaking and destabilizing effects~\citep{akbari2014catenoid, akbari2014bridge}.

Experimental studies on the statics of liquid bridges between equal circular discs are extensive. Using neutral-buoyancy experiments, \cite{sanz1983minimum} ascertained the minimum-volume stability limit in the slenderness range $0<\Lambda<6$. \cite{Russo1986154} determined the maximum-volume stability limit in a similar set-up, showing that axisymmetric liquid bridges non-axisymmetrically bulge when their interface is tangent to the discs. The experiments of \cite{slobozhanin1997bifurcation} provided further insights on this stability limit. Here, the stability limit corresponds to pitchfork bifurcations. Moreover, they showed that, above (below) the slenderness $\Lambda\simeq0.4946$, liquid bridges continuously (abruptly) bulge into a non-axisymmetric shape. Other studies considered the effect of gravity on the stability limits of axisymmetric~\citep{bezdenejnykh1992experimental} and non-axisymmetric liquid bridges~\citep{bezdenejnykh1999experimental} between equal discs. Emphasizing the destabilizing effect of gravity on nearly cylindrical liquid bridges, \cite{lowry1994stabilization} experimentally demonstrated that subjecting liquid bridges to an external laminar flow suppresses interfacial disturbances, thereby stretching the stability limit beyond that of static bridges. 

Surface imperfections (\eg, heterogeneity and roughness) complicate the equilibrium of gas-liquid-solid contact lines on real surfaces compared to ideal surfaces~\citep{joanny1984model, qian2011motion, chen2013modeling}. Contact lines remain pinned on real surfaces as long as the equilibrium contact angle is between the receding and advancing contact angles, and otherwise freely move~\citep{gao2006contact}. The receding contact angle is a key geometric parameter that affects the dispensed-drop size when the contact line is free~\citep{akbari2014bridge, qian2011motion}. \cite{chen2013modeling} experimentally and numerically studied the effect of contact-angle hysteresis on the the evolution and adhesion force of liquid bridges with two free contact lines. Similarly to \cite{qian2011motion}, experiments were conducted in a liquid-gas system where the gravity effect is alleviated by small bridge dimensions. Using similar experiments, \cite{chen2014liquid} examined the bridge breakup, showing that the liquid transfer ratio is correlated with the difference between the receding contact angles on the plates. 

When stretching liquid bridges, there is a critical length, identified with the static stability limit~\citep{qian2011motion}, where the dynamics dramatically change. Below this length, the dynamics are slow and the interface evolves quasi-statically. However, the dynamics are fast above the critical length where  fluid viscosity and inertia become significant as the interface nears pinch-off, and the liquid and interface velocities approach $u_{w}$. Because the dynamics in the pinch-off phase are much faster than in the quasi-static phase~\citep{eggers1997nonlinear}, the pinch-off length is commonly approximated as the static maximum-heigh stability limit in the literature~\citep{qian2011motion, chen2014liquid}.

Studying the relationship between the contact-line constraints and the stability limits of liquid bridges, \cite{akbari2014bridge} demonstrated that pinned-pinned liquid bridges can be stretch beyond the maximum-length stability limit of pinned-free bridges for the same liquid volume. Consequently, unstable states of the latter can be accessed during drop deposition if surface imperfections constrain the disturbances at the contact line. This has significant implications for the breakup dynamics and liquid transfer on (real) surfaces with contact-angle hysteresis. Depending on the surface wettability and drop volume, the contact angle may fall between or outside the receding and advancing contact angles when stretching liquid bridges. Hence, the bridge can undergo transitions from pinned-pinned to pinned-free contact lines (and vice versa) during stretching. These complications raise new, non-trivial questions as to how the stability limits and dispensed-drop volume are influenced by contact-angle hysteresis, and which stability limit (with respect to pinned-pinned or pinned-free disturbances) determines the breakup length.  

In this paper, we address the forgoing questions by studying pinned-pinned to pinned-free transitions and their respective stability limits during drop deposition on surfaces with contact-angle hysteresis. In particular, we show that, contrary to the common notion, there are cases where liquid bridges do not break at a critical state (\ie, the pinned-pinned or pinned-free stability limit). This observation has not been reported in the literature, as far as we are aware. Moreover, we experimentally verify the theoretical predictions of the maximum- and minimum-slenderness stability limits~\citep{akbari2014bridge} when the contact line is free at breakup. The stability limits of liquid bridges with pinned and moving contact lines are compared, demonstrating the destabilizing effect of free contact lines. To simulate zero gravity using the density matching technique, liquid bridges of silicon oil were formed in a water-methanol solution. The contact-angle effect was then studied by adding surfactant to the aqueous phase.

\section{Materials and methods} \label{sec4:materials}

\begin{figure} [t]
\centering
\includegraphics[scale=0.4]{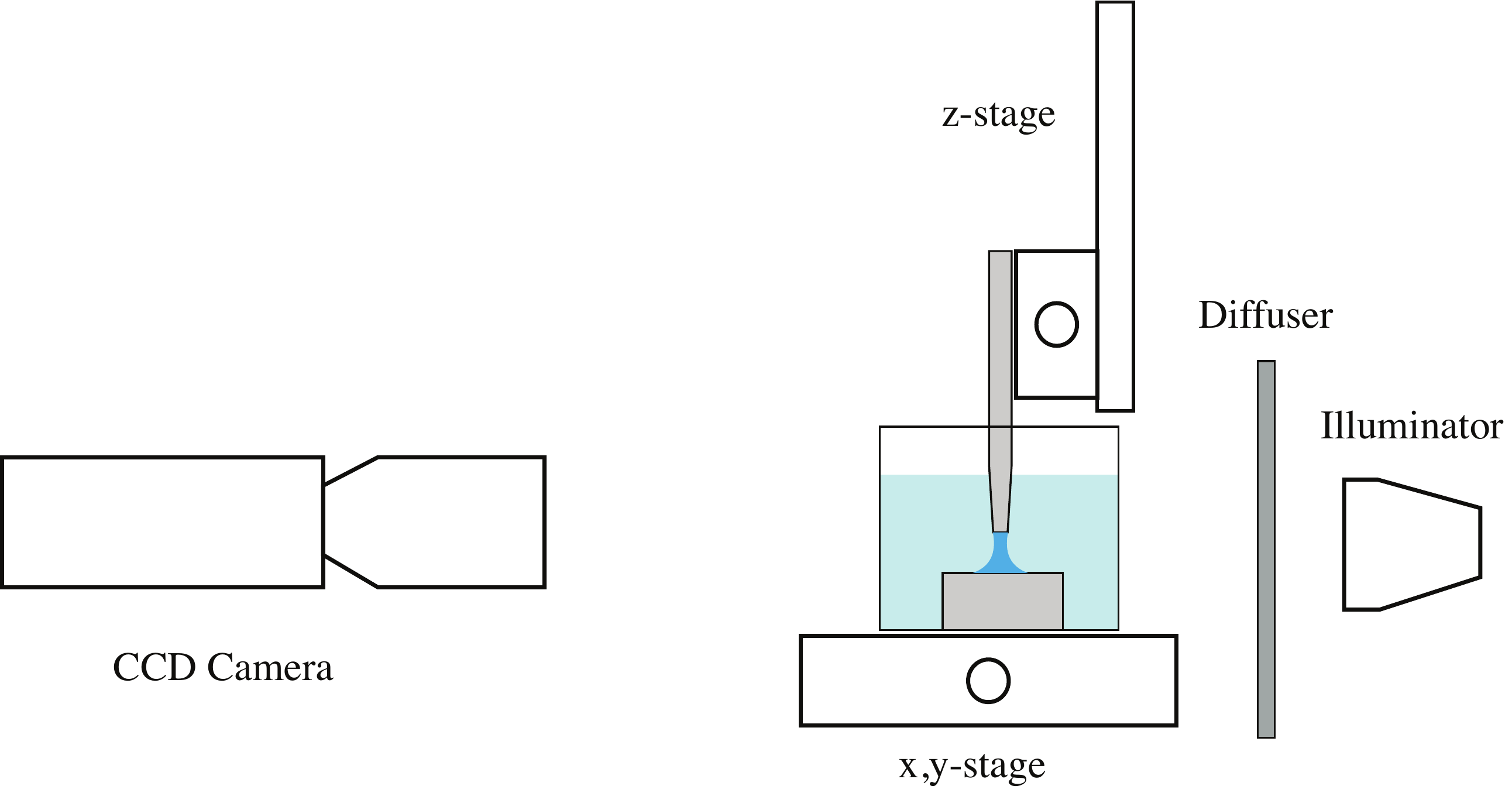}
 \caption{Schematic of the experimental setup.}
\label{fig4:figure1}
 \end{figure}

Experiments were performed in a cubic Plateau tank under neutrally buoyant conditions (Fig.~\ref{fig4:figure1}). Silicon oil (5 cSt, Sigma Aldrich) with specific gravity 0.92 was used as the bridge in a water-methanol solution (volumetric mixing ratio 42:58) bath. The composition of the bath solution was adjusted so that its density matched that of silicon oil at the experiment temperature ($\approx 20^{\circ}$C). Using a microsyringe, a drop with a prescribed volume in the range $5$--$50$~$\mu$l was deposited onto a plastic coverslip (Fischer Scientific), which had been soaked in a $0.1$~M hydrochloric acid solution, rinsed with DI water, and placed in the tank. A bridge was produced by gently pressing a needle with tip diameter $1.5$~mm into the drop. The needle was mounted on a one-dimensional vertical translation stage to control the bridge length, and the tank was placed on a two-dimensional positioning stage to align the drop and needle centers before contact, ensuring that the bridge is axisymmetric. The maximum (minimum) slenderness stability limit was ascertained by stretching (squeezing) the bridge until the bridge ruptured (bulged asymmetrically). A CCD camera (Prosilica GX1050, Allied Vision) with a $5\times$ lens (Nikon GMicro-NIKKOR) was used to record the bridge dynamics. Images were analyzed using an in-house Matlab script. The bridge contact angle with the coverslip was adjusted by changing the interfacial tensions in the system using the anionic surfactant sodium dodecyl sulfate (SDS) (Sigma Aldrich) at concentrations in the range $0$--$10$~g~l$^{-1}$.

\section{General behaviour of liquid bridges} \label{sec4:general}

\begin{figure} [t]
\centering
\includegraphics[scale=0.7]{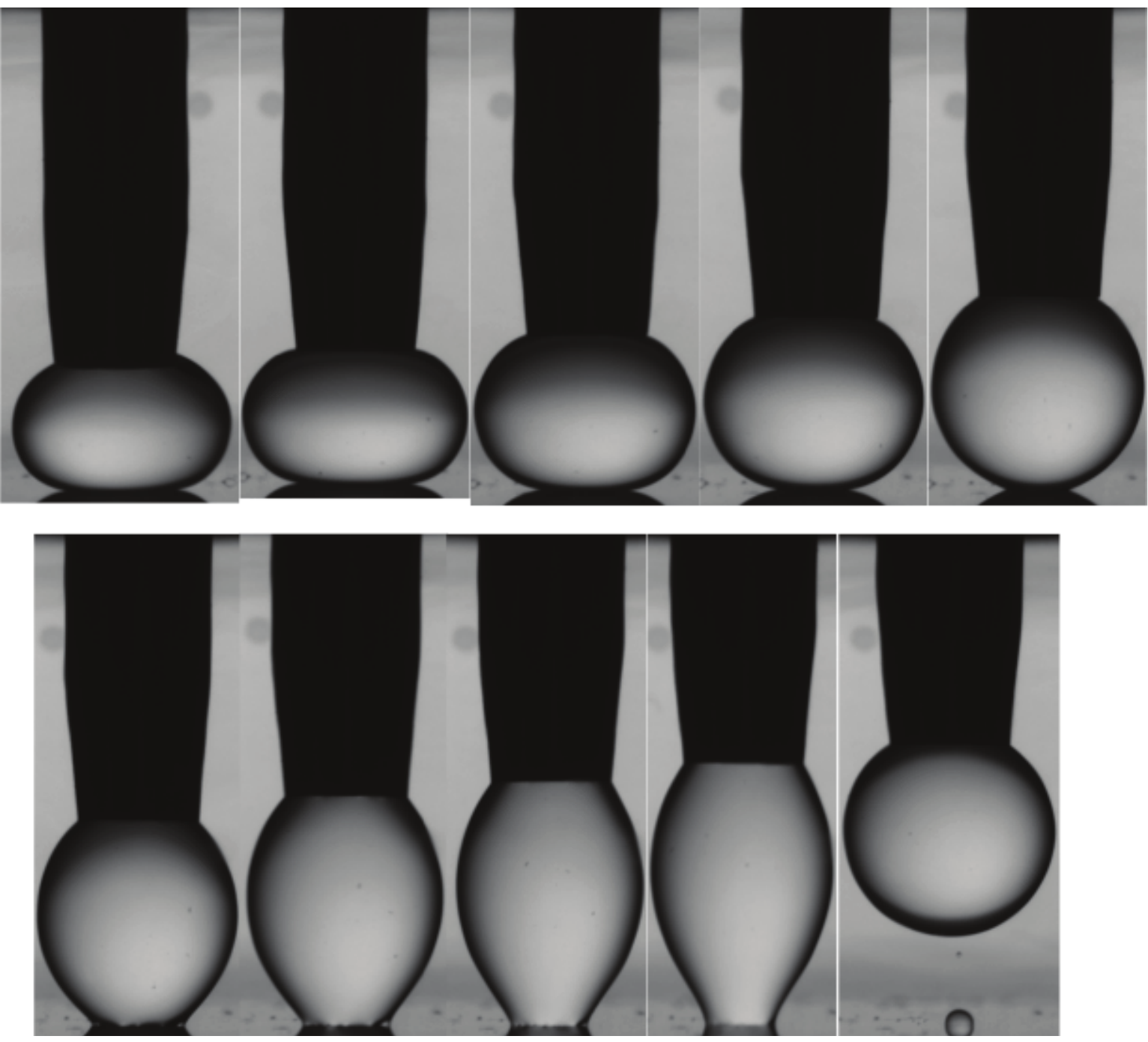}
 \caption{Representative stretching (bottom) and squeezing (top) sequences ($20$~$\mu$l drop).}
\label{fig4:figure2}
 \end{figure}

\begin{figure} 
\centering
\includegraphics[width=\linewidth]{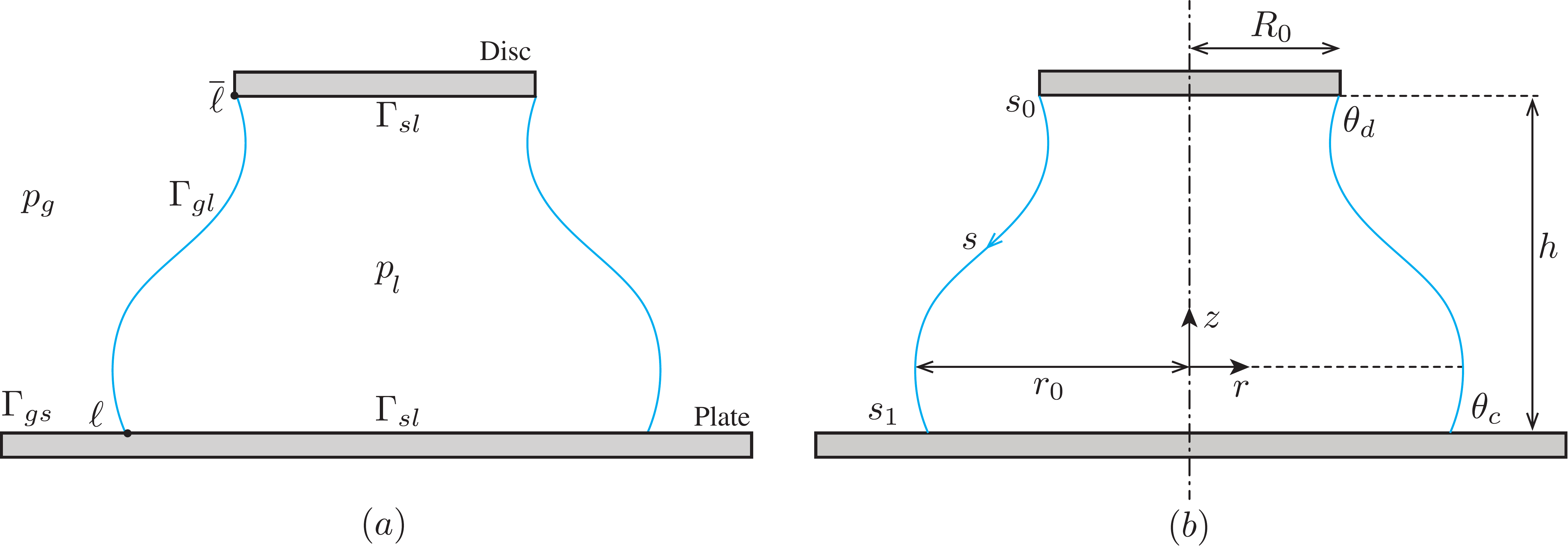}
 \caption{Weightless liquid bridge; (a) schematic and (b) coordinate system with meridian curve parametrization.}
\label{fig4:figure3}
 \end{figure}

Since the is was hollow with a sharp edged tip, bridges were always pinned to the needle. After a drop was deposited onto a coverslip, it was squeezed in $0.01$ inch steps to reach the minimum-slenderness stability limit, at which it bulged asymmetrically; bridges were imaged at each step. The maximum-slenderness stability limit was similarly measured by stretching bridges until rupture. Figure~\ref{fig4:figure2} shows a typical sequence during squeezing and stretching of a $20$~$\mu$l drop. In the rotund limit, non-axisymmetric drop deformations displace the contact line, leading to different results upon repeating the experiment. In the slender limit, the bridge breaks into two primary drops, leaving several satellite drops suspended in the bath. Without SDS, the contact line moved only at breakup for small drop volumes (less than $10$~$\mu$l), and was otherwise pinned at larger volumes. Since the emphasis in this work is on the role of moving contact lines, SDS was added to the bath to reduce the receding contact angle (see Fig.~\ref{fig4:figure3}) of the bridge on the coverslip. 

\section{Theory} \label{sec:theory}

Consider a liquid of volume $v$ bridging a circular disk with radius $R_{0}$ and a large plate. The disc and plate are separated by a distance $h$, as shown in Fig.~\ref{fig4:figure3}. The bridge is pinned to the disc and is free to slide horizontally on the plate. A small Bond number ($\mbox{Bo}\ll1$) is achieved by density matching, so the gravity force is negligible. Consequently, there is a constant pressure differential between the non-hydrostatic pressure of the bridge $p_{l}$ and the surrounding fluid $p_{g}$. Here, the surface tension between the phases $i$ and $j$ is denoted $\gamma_{ij}$ with $\Gamma_{ij}$ the corresponding interfacial surface area. The contact and dihedral angles that the interface $\Gamma_{gl}$ forms with the plate and disc are denoted $\theta_{c}$ and $\theta_{d}$, respectively. The cylindrical volume $V=v/(\pi R_{0}^{2}h)$, scaled volume $v^{*}=v/(4\pi R_{0}^{3}/3)$, scaled pressure (mean curvature) $Q=qR_{0}$, and slenderness $\Lambda=h/R_{0}$ are the dimensionless parameters with which the liquid bridges are specified. Note that $q=(p_{g}-p_{l})/\gamma_{gl}$ measures the non-hydrostatic pressure differential~\citep{myshkis1987low}. Solving the Young-Laplace equation furnishes the equilibrium meridian curve 
\begin{equation}
    	\left\{ \begin{array} {l}
	\rho(\tau)=\sqrt{1+a^{2}+2a \cos \tau}\\
	\xi(\tau)=\int_{0}^{\tau} \frac{1+a\cos t}{\rho(t)} \mbox{d}t
	\end{array} \right.,
	 \label{eqn:eq1}
\end{equation}
where $a=\rho(0)-1$ and $\tau$ is the mean-curvature-scaled arclength~\citep{akbari2014bridge}. The scaled arclength at the hole edge $\bar{\ell}$ and the contact line $\ell$ are denoted $\tau_{0}$ and $\tau_{1}$, respectively. \cite{akbari2014bridge} showed that critical surfaces at the minimum-slenderness stability limit, corresponding to the upper boundary of the stability region, are nodoids with $\theta_d=0$, which are well approximated by
\begin{equation}
\centering
\begin{split}
 	V &=1+\frac{1}{4} \sec^{4}(\theta_{c}/2)(\pi-\theta_{c}+\cos\theta_{c} \sin\theta_{c})\Lambda\\
	&-\frac{1}{384}\sec^{8}(\theta_{c}/2) [-97+24(\pi-\theta_{c})^{2}-136\cos\theta_{c}-32 \cos(2\theta_{c})\\
	&+8\cos(3\theta_{c}) +\cos(4\theta_{c}) +24(\pi-\theta_{c})\sin(2\theta_{c})]\Lambda^{2}+O(\Lambda^{3})
\end{split}
	 \label{eqn:eq2}
\end{equation}
in $V$ versus $\Lambda$ stability diagrams. 

Note that pinned-free and pinned-pinned liquid bridges are, respectively, specified by $\textbf{p}=(\Lambda,V,\theta_{c})$ and $(\Lambda,V,K)$, where $K$ is the ratio of the lower to upper contact-line radii. When stretching the bridge, $K$ varies at fixed $\theta_{c}$ in pinned-free bridges, and $\theta_{c}$ varies at fixed $K$ in pinned-pinned bridges. Furthermore, we determine stability along equilibrium branches using Myshkis's variational method, detailed by \cite{akbari2014bridge}.

 \begin{figure} [t]
\centering
\includegraphics[scale=0.5]{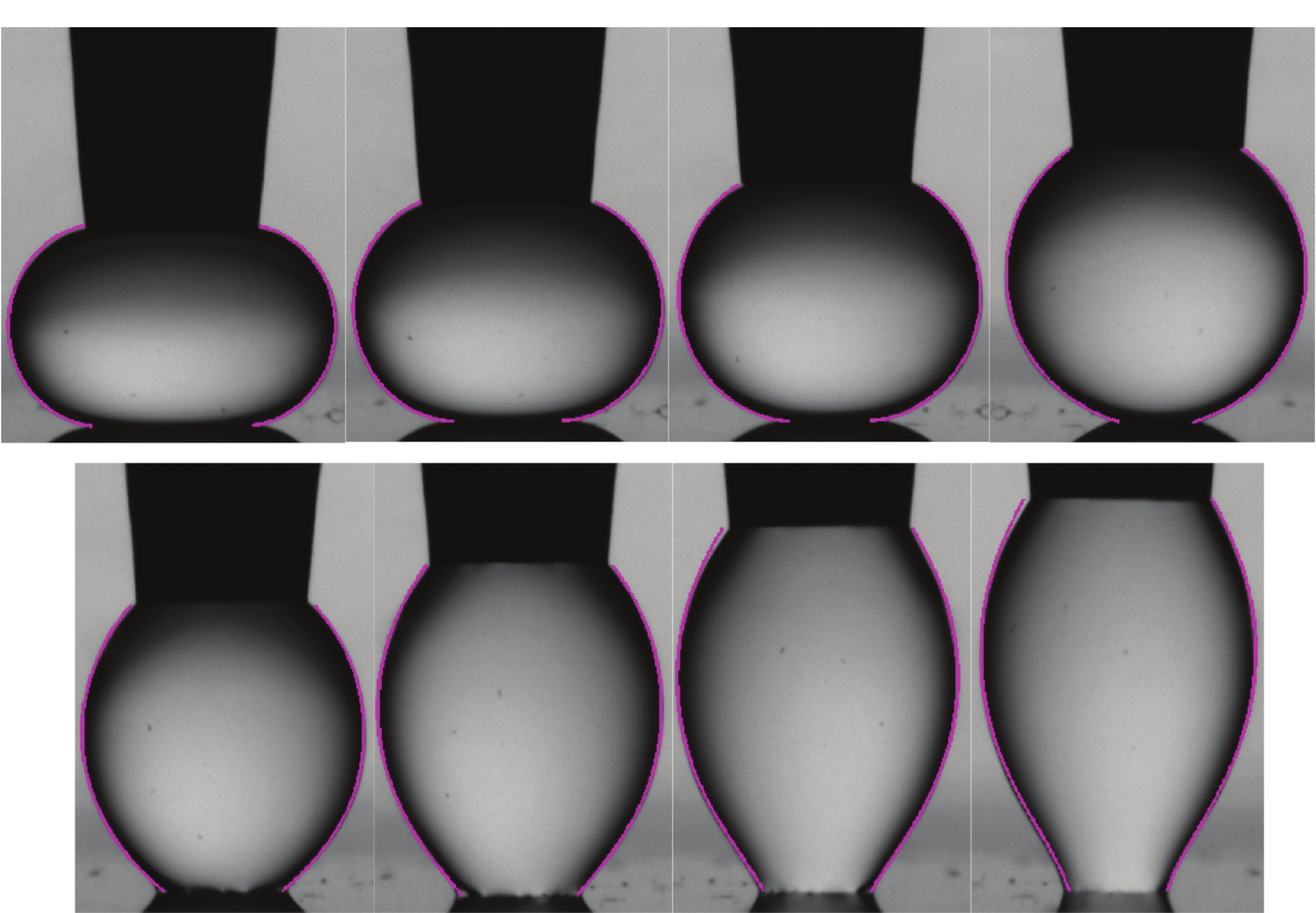}
 \caption{Fitting theoretical meridian curves to bridges boundaries in stretching (bottom) and squeezing (top) sequences ($20$~$\mu$l drop).}
\label{fig4:figure4}
 \end{figure}
 
\section{Feature extraction} \label{sec4:feature}
The bridge interface with the bath solution was extracted using a gradient-based edge detection method with a Gaussian optimal smoothing filter~\citep{marr1980theory}. In this method, pixels on an interface are identified by finding maxima in the first directional derivative of intensity; or, equivalently, seeking zero-crossings in the second directional derivative. Derivatives were taken along normals to interfaces using high-order (8-10 points) central schemes. Then, the analytical solution of the bridge meridian curve, given by Eq.~(\ref{eqn:eq1}), was fitted to the extracted interfaces. Here, the unknown parameters $(Q_{c},\tau_{0},\tau_{1},a)$ were determined by minimizing the root-mean-squared normal distances between the extracted interface pixels and the theoretical meridian curve. Figure~\ref{fig4:figure4} shows typical results of the image-processing script in stretching and squeezing experiments.    

 \begin{figure} [t]
\centering
\includegraphics[width=\linewidth]{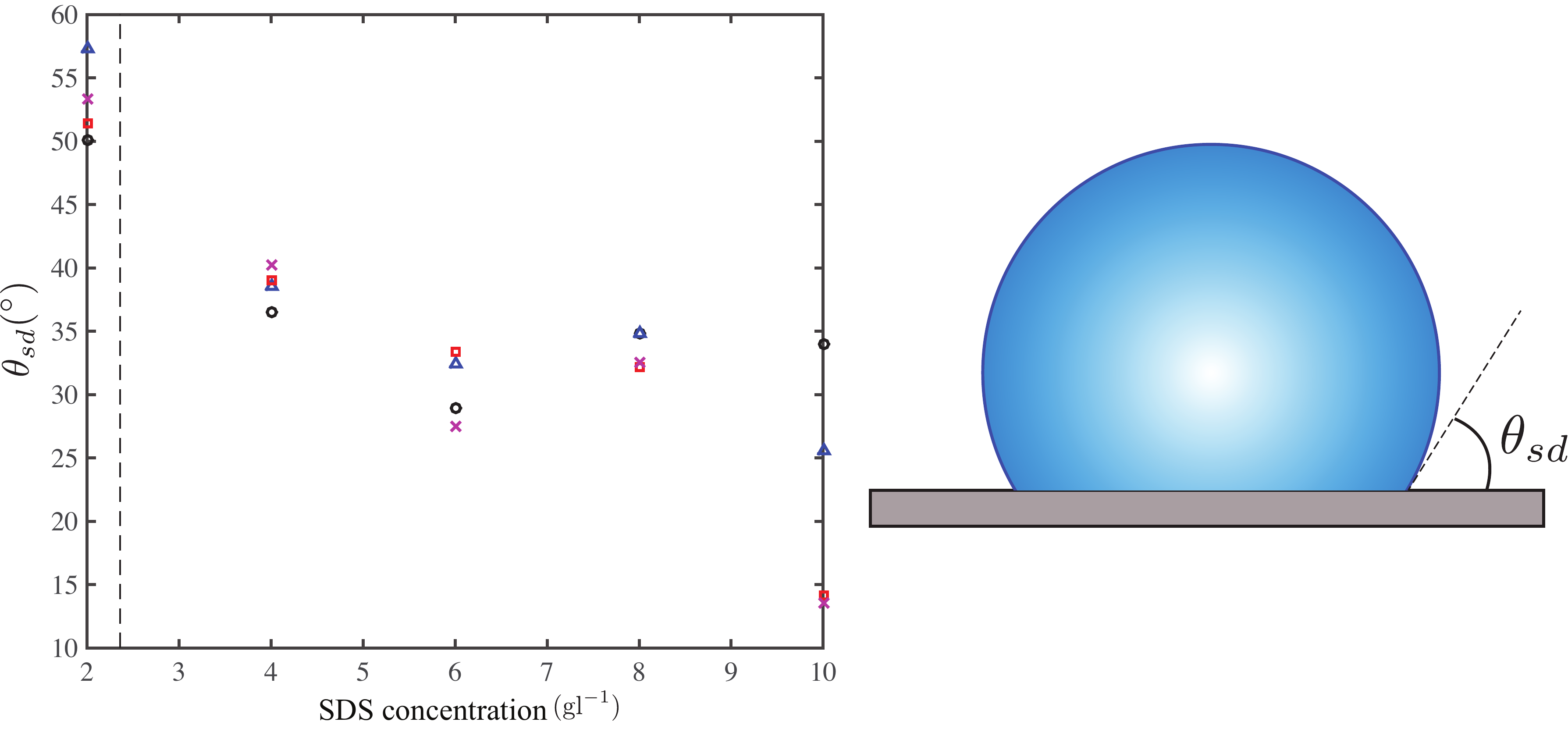}
 \caption{The surfactant-concentration effect on the sessile-drop contact angle $\theta_{sd}$ at drop volumes $5$~$\mu$l (${\scriptstyle \bigcirc}$), 10~$\mu$l ($\triangle$), 1$5$~$\mu$l ($\Box$), $20$~$\mu$l ($\times$). Dashed line indicates the critical micelle concentration of the surfactant in pure water at 25$^{\circ}$C~\citep{mukerjee1971critical}.}
\label{fig4:figure5}
 \end{figure}
 
\section{Results and discussion} \label{sec4:results}
\subsection{Surfactant effect} \label{sec4:surfactant}
As previously stated, for large drops, the contact angle $\theta_{c}$ remains smaller than the receding contact angle $\theta_{r}$ during stretching. Thus, the contact line is pinned to the coverslip at breakup. To assess the stability limits of liquid bridges with a free contact line over a wider range of drop volumes, the contact angle was reduced by adding SDS to the bath solution. The critical micelle concentration (CMC) of SDS in pure water at $25^{\circ}$C is $\approx 2.36$~g~l$^{-1}$~\citep{mukerjee1971critical}. Previous measurements in the literature have shown that the air-water surface tension exhibits no minimum near the CMC~\citep{lucassen1981surface}. According to Young's equation~\citep{myshkis1987low}, this implies that the contact angle also does not exhibit a minimum if the air-solid and water-solid surface tensions remain constant. Furthermore, previous reports in the literature indicate that, unlike the advancing contact angle, the receding contact angle can be sensitive to the drop volume~\citep{drelich1996contact}. Therefore, we examine how the contact angle varies with the SDS concentration at various drop volumes, in the silicon oil-water-methanol system, to determine the surfactant concentration at which the contact angle is minimum for all volumes. The contact angle was measured using the sessile-drop method~\citep{bachmann2000development}. 
 
 \begin{figure}[t] 
\centering
\includegraphics[width=\linewidth]{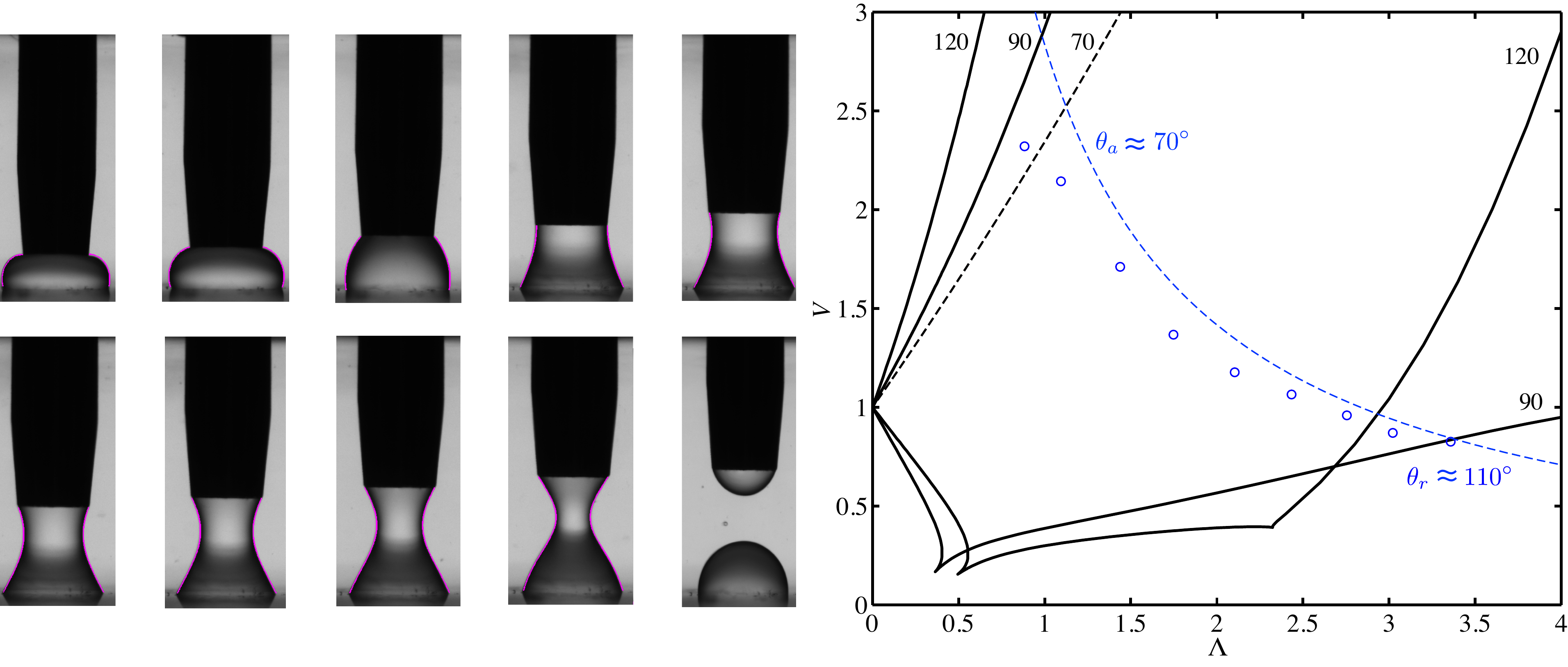}
 \caption{Comparison of the theoretical prediction and experimental measurement of the stability limits with drop volume $v\approx 5~\mu$l, receding contact angle $\theta_{r}\approx110^{\circ}$, advancing contact angle $\theta_{a}\approx70^{\circ}$, and without surfactant (right).  An image sequence of the bridge evolution corresponding to the data points (left). Dashed blue and black lines respectively indicate the constant-$v$ isocontour at the dispensed drop volume and the maximum-volume stability limit estimated by Eq.~(\ref{eqn:eq2}) at $\theta_{c}=70^{\circ}$. Labels denote the contact angle in degrees.}
\label{fig4:figure6}
 \end{figure}
 
 Figure~\ref{fig4:figure5} shows the surfactant effect on the contact angle. The contact angle decreases almost linearly around the CMC and below $\sim6$~g~l$^{-1}$ for all drop volumes. At higher concentrations, the relationship is nonlinear. Nevertheless, the smallest value of the contact angle in the range $2$--$10$~g~l$^{-1}$occurs at $10$~g~l$^{-1}$ for all volumes, except $5$~$\mu$l. Moreover, the contact angle for larger drops is affected more by the surfactant at this concentration. Since larger drops tend to have a pinned contact line at breakup more often than smaller drops, only the stability-limit results for experiments where the SDS concentration is $10$~g~l$^{-1}$ are reported for all drop volumes. At this concentration, the bridge contact line with the coverslip moved at all drop volumes.  
 
\subsection{Stability limits} \label{sec4:stability}

 The maximum-slenderness stability limit was determined by stepwise quasi-static stretching of a liquid bridge with fixed volume. Similarly, the minimum-slenderness stability limit was determined by stepwise quasi-static squeezing of a bridge. In all experiments, the contact line was either pinned or receding when stretching, and always advancing when squeezing. Therefore, when the contact line is moving, the receding contact angle $\theta_{r}$ is the relevant contact angle in stretching, and the advancing contact angle $\theta_{a}$ is the relevant contact angle in squeezing. Figure~\ref{fig4:figure6} shows an image sequence during stretching and squeezing of a $5~\mu$l drop. Here, no surfactant was added to the bath, and the contact line moved during stretching and squeezing. The receding contact angle at breakup was measured $\theta_{r}\approx110^{\circ}$; thus, the corresponding data point (far right) in the stability diagram is expected to fall between the lower boundary of the stability region~\citep{akbari2014bridge} for $\theta_{c}=90^{\circ}$ and $120^{\circ}$, as demonstrated in Fig.~\ref{fig4:figure6}. The advancing contact angle was measured $\theta_{a} \approx 70^{\circ}$; here, the minimum-slenderness stability limit is estimated by Eq.~(\ref{eqn:eq2}) and then compared with the measured value. As shown in Fig.~\ref{fig4:figure6}, experimental measurements of the stability limits are in good agreement with the theoretical predictions of \cite{akbari2014bridge}. Note that, because the needle is hollow, part of the initial drop volume is driven into the needle in squeezing experiments, so the bridge volume does not reflect the initial drop volume; moreover, data points deviate more from the constant-$v$ isocontour corresponding to the initial dispensed volume (dashed blue line) near the upper boundary of the stability region. 
 
\begin{figure} [h]
\centering
\includegraphics[width=\linewidth]{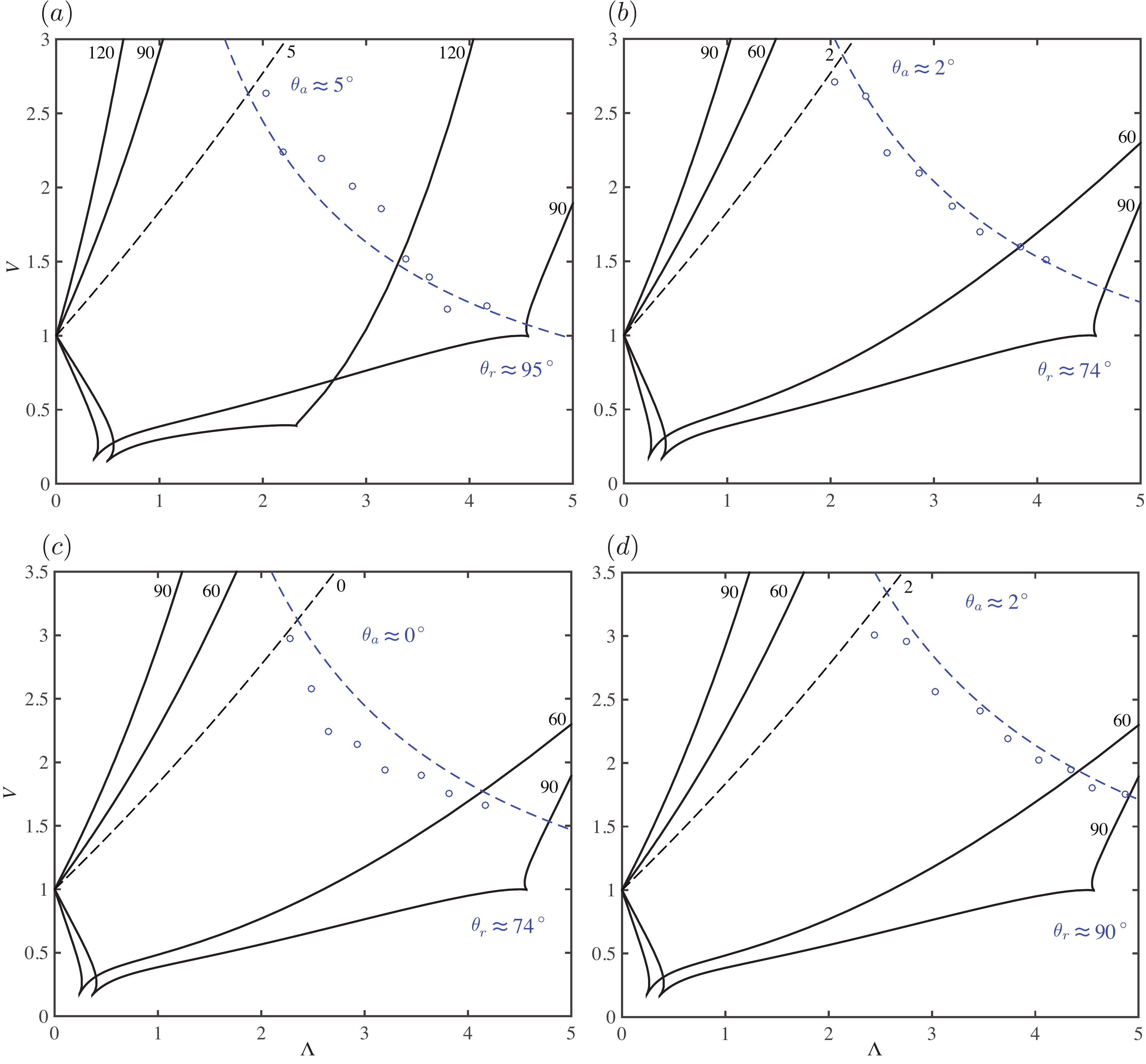}
 \caption{Same as Fig.~\ref{fig4:figure6}, but with SDS concentration $10$~g~l$^{-1}$ at drop volumes (a) $v=10~\mu$l, (b) $v=12.5~\mu$l, (c) $v=15~\mu$l, and (d) $v=17.5~\mu$l.}
\label{fig4:figure7}
 \end{figure}

Stretching and squeezing experiments were conducted in the range $v=$~5--$20$~$\mu$l with SDS added to the bath. At $10$~g~l$^{-1}$ SDS, the advancing and receding contact angles drop to $\theta_{a}\approx$~0--5$^{\circ}$ and $\theta_{r}\approx$~75--95$^{\circ}$. Reasonable agreement is observed between the experimental measurements of the stability limits and theoretical predictions of \cite{akbari2014bridge} (see Fig.~\ref{fig4:figure7}). 

 \begin{figure}
 \centering
\includegraphics[width=\linewidth]{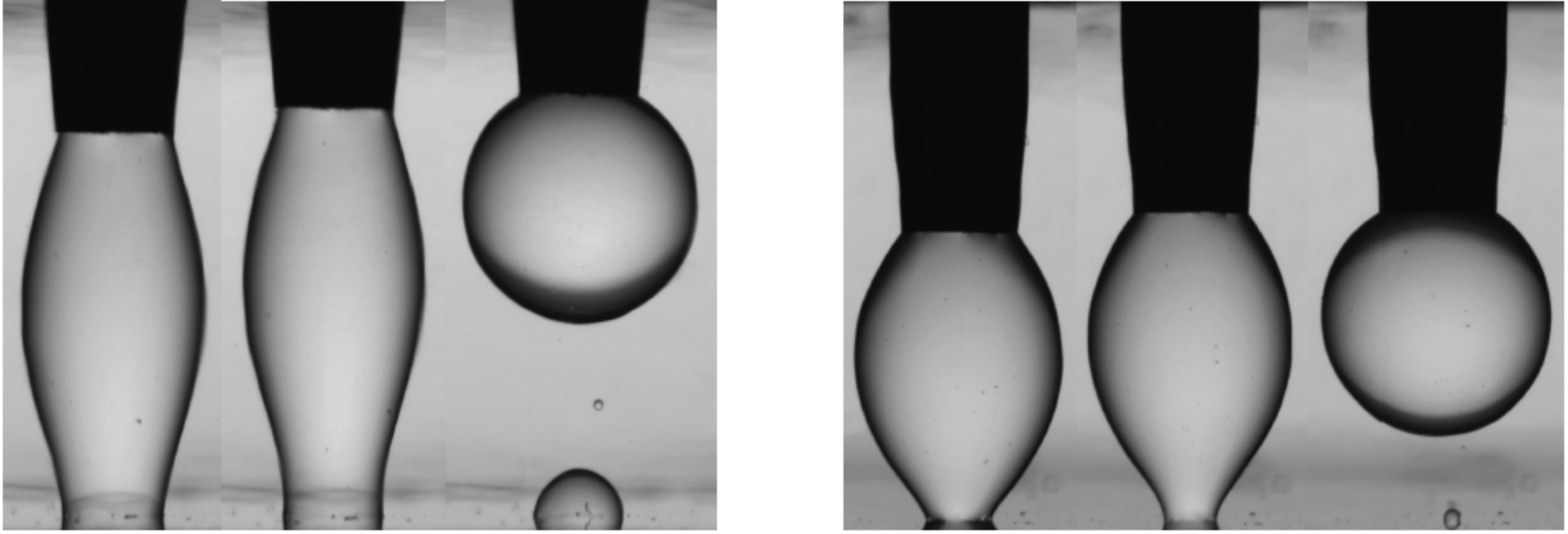}
 \caption{Contact-line effect on the breakup length of liquid bridges. Stretching a $20~\mu$l drop with $10$~g~l$^{-1}$ SDS in the bath, producing a pinned contact line with $\theta_{c}\approx84^{\circ}$ (left) and a free contact line with $\theta_{c}\approx81^{\circ}$ (right) at breakup.}
\label{fig4:figure8}
 \end{figure}
 
Figure~\ref{fig4:figure8} shows the effect of a free contact line on the maximum-slenderness stability limit. Here, a stretching experiment was conducted on two coverslips using a $20~\mu$l drop with $10$~g~l$^{-1}$ SDS. These coverslips exhibited slightly different contact angles at breakup, presumably due to different surface characteristics, so that $\theta_{c}$ was below $\theta_{r}$ at breakup on one (Fig.~\ref{fig4:figure8}, left panel) and $\theta_{c}$ reached $\theta_{r}$ before breakup on the other (Fig.~\ref{fig4:figure8}, right panel); consequently, the contact line was pinned on the former and free on the latter, and the slendernesses at breakup were measured $\Lambda_{b}\approx5.99$ and $\Lambda_{b}\approx4.94$, respectively. This $\approx20$\% decrease in the breakup length reflects the destabilizing effect of a free contact line, as theoretically predicted for static catenoidal and cylindrical liquid bridges~\citep{akbari2014catenoid}.

\begin{figure}
\centering
\includegraphics[width=\linewidth]{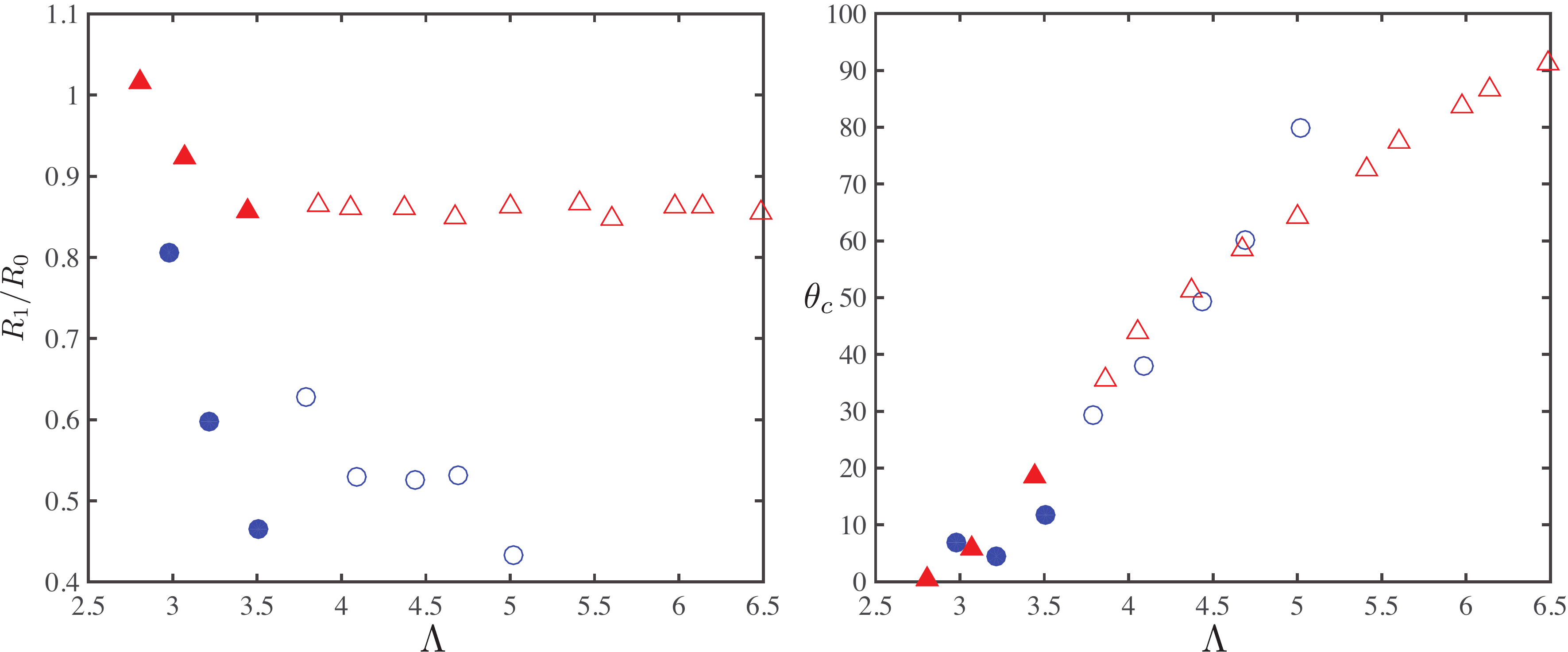}
 \caption{Squeezing (filled markers) and stretching (open markers) of a $20$~$\mu$l drop with $10$~g~l$^{-1}$ SDS, corresponding to the experiments shown in Fig.~\ref{fig4:figure8}. The radius of the meniscus contact line $R_1$ (left) and contact angle $\theta_{c}$ (right) versus slenderness are plotted when the bridge contact-line on the coverslip at breakup is free (${\scriptstyle \bigcirc}$, right panel in Fig.~\ref{fig4:figure8}) and pinned ($\triangle$, left panel in Fig.~\ref{fig4:figure8}).}
\label{fig4:figure9}
 \end{figure}
 
Figure~\ref{fig4:figure9} shows the radius of the meniscus contact line $R_1$ and contact angle $\theta_{c}$ in the experiments depicted in Fig.~\ref{fig4:figure8}. For the liquid bridge shown in the left panel of Fig.~\ref{fig4:figure8}, the contact line is pinned, and the contact angle varies during stretching. Here, the contact angle remains below the receding contact angle during the entire experiment. By contrast, for the liquid bridge shown in the right panel, the contact angle reaches the receding contact angle before breakup, and the contact line recedes during stretching. Here, the contact-line motion when stretching the bridge is significantly larger than in the left panel, indicating that the contact line on the coverslip shown in the left panel of Fig.~\ref{fig4:figure8} is more constrained than that in the right panel. 

Figure~\ref{fig4:figure9} also illustrates a qualitative difference between systems with and without contact-angle hysteresis. Advancing and receding experiments were performed on separate coverslips for the bridge with pinned contact lines (triangles). Therefore, the bridge does not experience hysteresis, and the contact line and contact angle vary continuously from stretching to squeezing. By contrast, squeezing and stretching were consecutively performed on the same coverslip for the bridge with a free contact line (circles). Here, the contact line retreats on a surface that is covered by the silicon oil during squeezing, which affects $\gamma_{sg}$ as the bridge is stretched. Consequently, the receding contact angle is different at a given contact-line position during stretching and squeezing, so that the contact line radius follows a different path when the bridge is stretched to its initial length and beyond. 

\subsection{Contact-angle hysteresis effect} \label{sec4:hysteresis}

\begin{figure} 
\centering
\includegraphics[width=\linewidth]{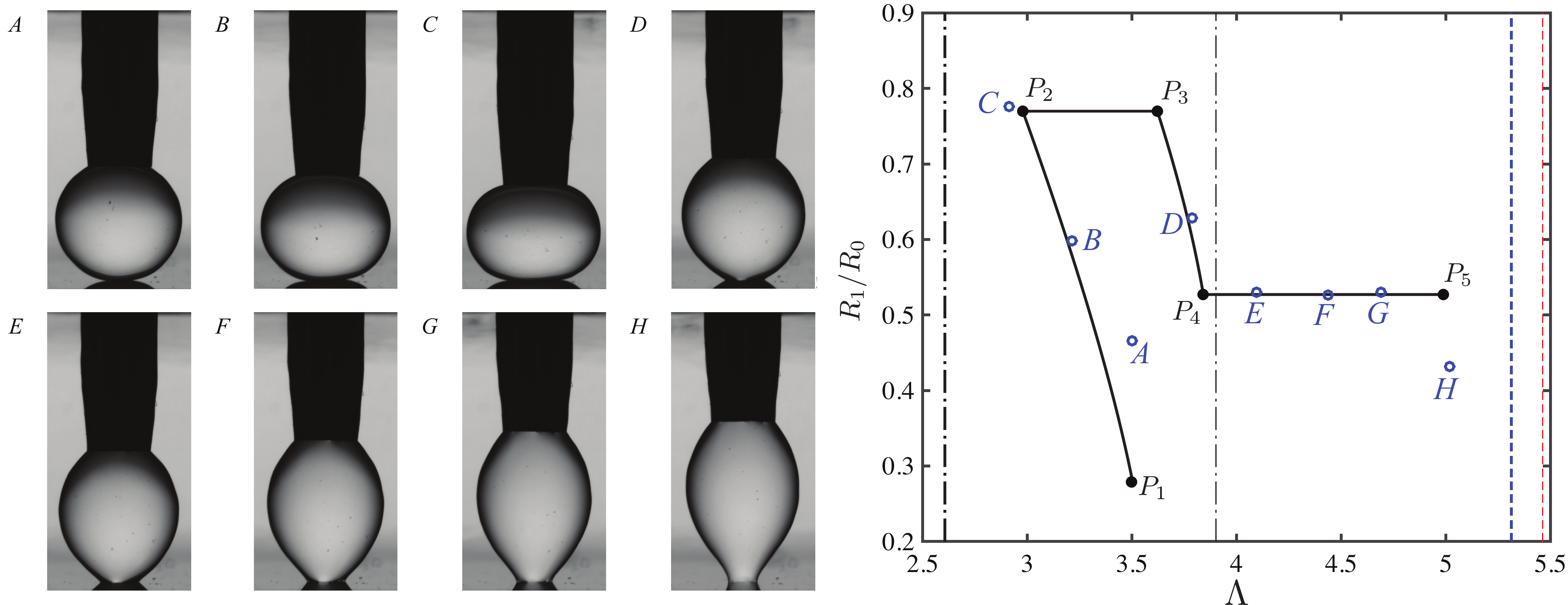}
 \caption{Comparison of experimental measurements (open circles) and theoretical predictions (solid lines) of the contact-line radius during the stretching and squeezing of a $20$~$\mu$l drop ($v^{*}\approx7.171$) on a substrate with an advancing ($\theta_{a}\approx3^{\circ}$) and two receding ($\theta_{r1}\approx30^{\circ}$, $\theta_{r2}\approx75^{\circ}$) contact angle(s). Vertical lines indicate the minimum-high stability limit at $\theta_{c}=3^{\circ}$ (thick dashed-dotted) and maximum-high stability limits at $\theta_{c}=30^{\circ}$ (thin dashed-dotted), $\theta_{c}=75^{\circ}$ (thick dashed), and $K=0.5271$ (thin dashed), where the contact line is free for the first three and pinned for the last. Equilibrium states are computed at fixed contact angle along $P_1P_2$ ($\theta_{c}=5^{\circ}$) and $P_3P_4$ ($\theta_{c}=30^{\circ}$), and at fixed contact-line radius along $P_2P_3$ ($K=0.7699$) and $P_4P_5$ ($K=0.5271$).}
\label{fig4:figure10}
 \end{figure}

In this section, we focus on data for the bridge with a free contact line at breakup (circles) in Fig.~\ref{fig4:figure9} and examine how the breakup length is associated with the maximum-length stability limit with respect to pinned-pinned and pinned-free disturbances. Figure~\ref{fig4:figure10} shows the bridge-evolution images, comparing the theoretical prediction of the contact-line radius and measured values. Reasonable agreement is observed between the measurements and theoretical predictions. The last image $H$ was recorded between stability loss and breakup. Here, the coverslip exhibited an advancing contact angle $\theta_{a}\approx3^{\circ}$ and two distinct receding contact angles $\theta_{r1}\approx30^{\circ}$ and $\theta_{r2}\approx75^{\circ}$. As previously stated, this can be attributed to changes in the coverslip interfacial tension with the bath solution upon retreating the contact line on a surface that is already covered by silicon oil. 

The contact line was free during the entire squeezing experiment, and the contact angle remained almost fixed at $\theta_{c}=\theta_{a}$. However, during stretching, the contact line was pinned ($K\approx0.7699$) when $\theta_{a}<\theta_{c}<\theta_{r1}$, free while the contact angle was almost fixed at $\theta_{c}=\theta_{r1}$, and pinned ($K\approx0.5271$) when $\theta_{r1}<\theta_{c}<\theta_{r2}$. The trajectory $P_1P_2$ indicates the equilibrium solution corresponding to the squeezing, whereas $P_2P_3$, $P_3P_4$, and $P_4P_5$ correspond to the foregoing pinned-pinned, pinned-free, and pinned-pinned phases of the stretching experiment. The point $P_5$ corresponds to a state where $\theta_{c}=\theta_{r2}$, indicating a transition from pinned-pinned bridges at $K=0.5271$ to pinned-free bridges at $\theta_{c}=75^{\circ}$. Pinned-free bridges along $P_3P_4$ are not stretched beyond their stability limit (thin dashed-dotted line), niether are pinned-pinned bridges along $P_4P_5$ beyond theirs (thin dashed line). Note that the slenderness at $P_5$ is also well below the pinned-free stability limit (thick dashed line) where the foregoing transition occurs. However, experimental data indicate that the bridge loses stability during the transition, and it does not correspond to the pinned-pinned or pinned-free stability limit. 

\begin{figure} 
\centering
\includegraphics[width=\linewidth]{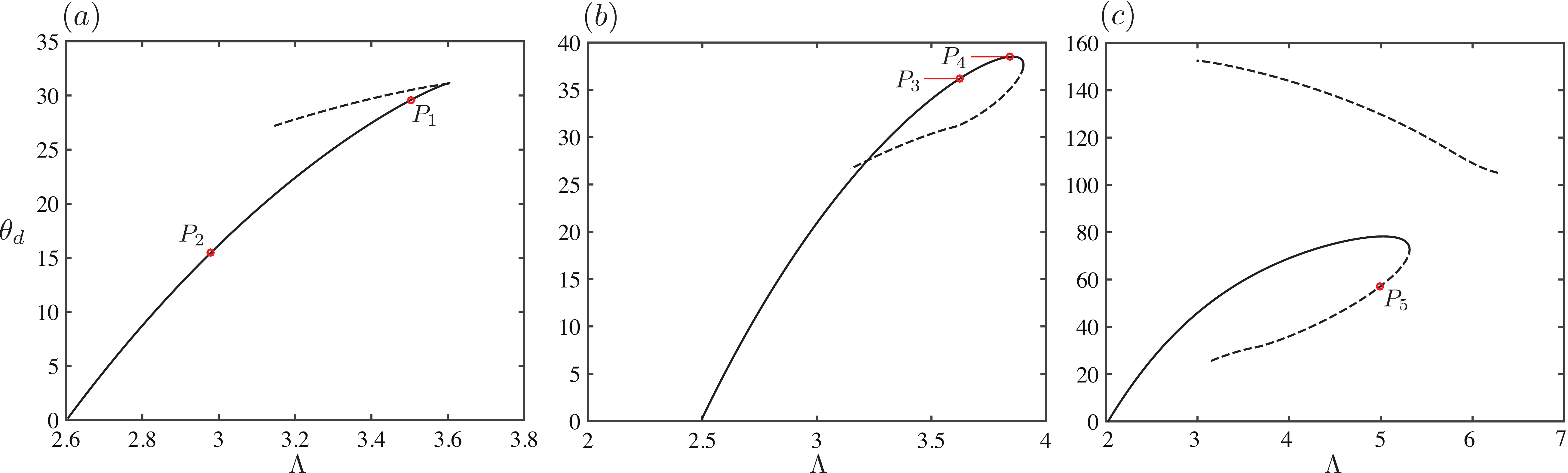}
 \caption{Free-contact line equilibrium branches of a fixed-volume ($v^{*}=7.171$) liquid bridge, indicating stable (solid) and unstable (dashed) states at (a) $\theta_{c}=3^{\circ}$, (b) $\theta_{c}=30^{\circ}$, and (c) $\theta_{c}=75^{\circ}$. The terminal points $P_{1-5}$ are shown to identify the stability of equilibrium trajectories in Fig.~\ref{fig4:figure10}.}
\label{fig4:figure11}
 \end{figure}
 
 \begin{figure} 
\centering
\includegraphics[width=\linewidth]{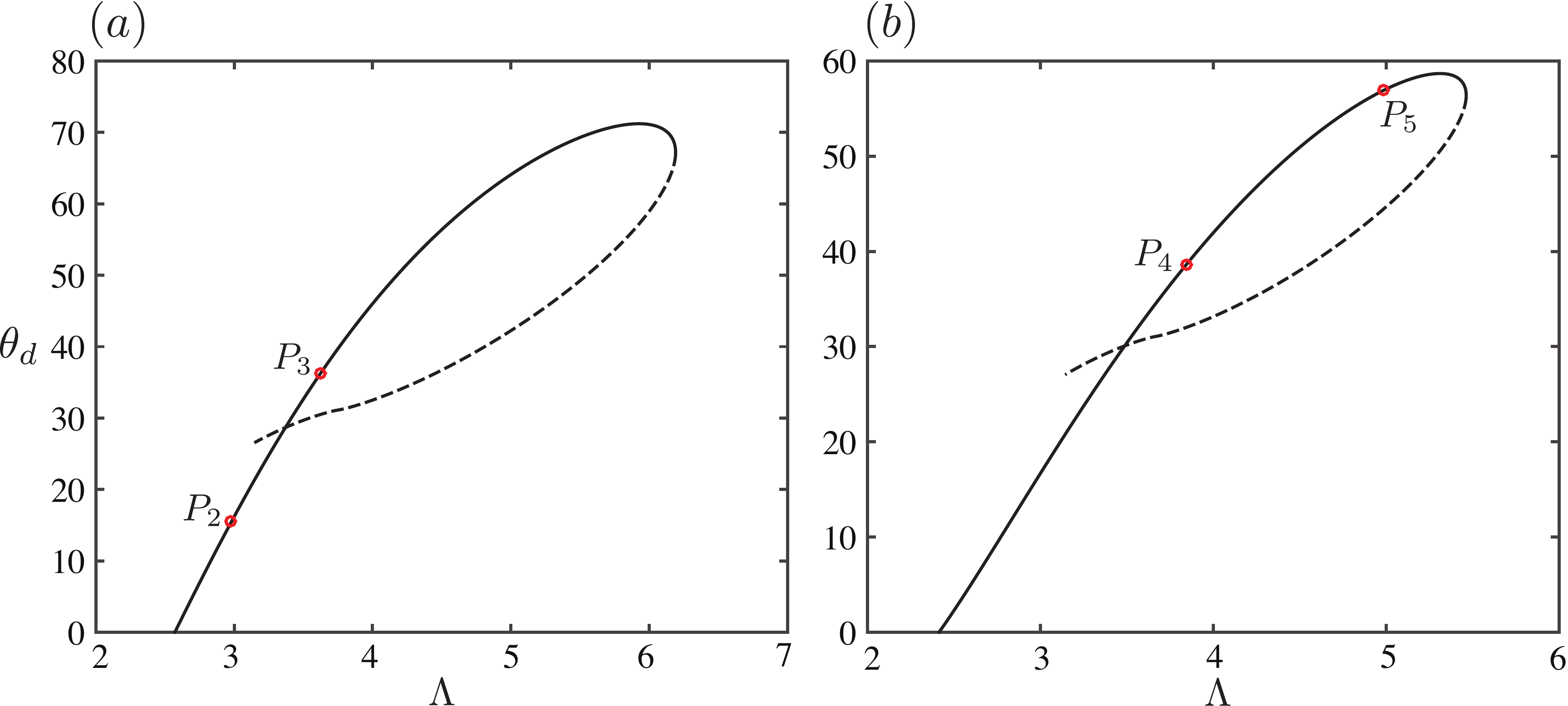}
 \caption{Pinned-contact line equilibrium branches of a fixed-volume ($v^{*}=7.171$) liquid bridge, indicating stable (solid) and unstable (dashed) states at (a) $K=0.7699$ and (b) $K=0.5271$. The terminal points $P_{2-5}$ are shown to identify the stability of equilibrium trajectories in Fig.~\ref{fig4:figure10}.}
\label{fig4:figure12}
 \end{figure}
 
 To understand why stability is not lost at a critical state, we locate the stretching and squeezing trajectories of Fig.~\ref{fig4:figure10} on equilibrium branches in Figs.~\ref{fig4:figure11} and \ref{fig4:figure12}. Here, the bridge follows stable equilibrium branches along the entire $P_1P_5$ trajectory and before the transition at $P_5$. We note that the bridge lies on a stable branch with respect to pinned-pinned disturbance (see Fig.~\ref{fig4:figure12}b) before the turing point. Upon transition at $\theta_{c}=\theta_{r2}$, the bridge is exposed to pinned-free disturbances ($P_5$ in Fig.~\ref{fig4:figure11}c). Although the slenderness is smaller than the maximum-slenderness stability limit (at the turning point), this state lies on an unstable branch, where the bridge loses stability to pinned-free perturbations. This implies that, it is possible for liquid bridges to break during the pinned-pinned to pinned-free transition rather than a critical state at a turning point in systems with contact-angle hysteresis; therefore, \emph{the breakup length is not always associated with the static maximum-length stability limit.}   
 
These observations also have significant implications for the breakup dynamics and dispensed-drop volume. Recall, pinned-pinned liquid bridges are more stable than pinned-free ones. Consequently, the foregoing transition provides access to unstable states of pinned-free bridges, far from critical sates. Before the transition, the contact line is pinned, and, depending on the bridge volume and receding contact angle, can be stretched to a state that is highly unstable to pinned-free perturbations. Once the receding contact angle is reached, the constraint at the contact line is relaxed, and the interface is exposed to a larger set of perturbations. This leads to a dramatic stability loss at a point that does not coincide with a critical state ($P_5$ in Figs.~\ref{fig4:figure12}b and \ref{fig4:figure11}c). Note that the energy barrier (potential well) disappears at critical states, and the instability margin\footnote{The notion of the instability margin in this paper is the stability-margin counterpart of \citet{slobozhanin2002stability} for unstable bridges. Note that liquid bridges break into several primary and satellite drops upon stability loss at the maximum-length stability limit~\citep{meseguer1995review}. Thus, the chain of drops arising upon breakup is the most stable (having the deepest potential well) state that is dynamically accessible to unstable bridges. Accordingly, the instability margin is defined as the potential-energy difference between the most stable and unstable states.} generally grows as an unstable state moves farther away from its critical state along the respective equilibrium branch~\citep{Myshkis1965on, slobozhanin2002stability}. Potential energy of the instability margin can be transformed to kinetic energy upon stability loss, acting as the deriving force for the near-singularity dynamics~\citep{eggers1997nonlinear}. 

An appreciable difference between the breakup dynamics upon stability loss at a critical state and at an unstable state away from its critical state is expected. At a critical state (\eg, turning point), the energy barrier disappears in the direction of the critical perturbation, so the interface accelerates in the same direction with a kinetic energy that is proportional to the disturbance magnitude. However, at an unstable state away from its critical state, the instability margin significantly amplifies the critical perturbation, leading to more dramatic dynamics and a significant impact on the dispensed-drop size. As previously stated, \cite{qian2009micron} reported small drop sizes in a pressure-controlled deposition due to fast dynamics near the contact line. Here, achieving a fast-receding contact line is assisted by the withdrawal of the liquid near the plate, which is less significant in the volume-controlled case~\citep{qian2011motion}. However, breakup at an unstable state with a large instability margin can greatly influence the dynamics in volume-controlled deposition, potentially favouring smaller deposited drops. Further studies are required to examine the possibility of stability loss at unstable states with large instability margins, and their ensuing dynamics. 

\begin{figure} 
\centering
\includegraphics[scale=0.4]{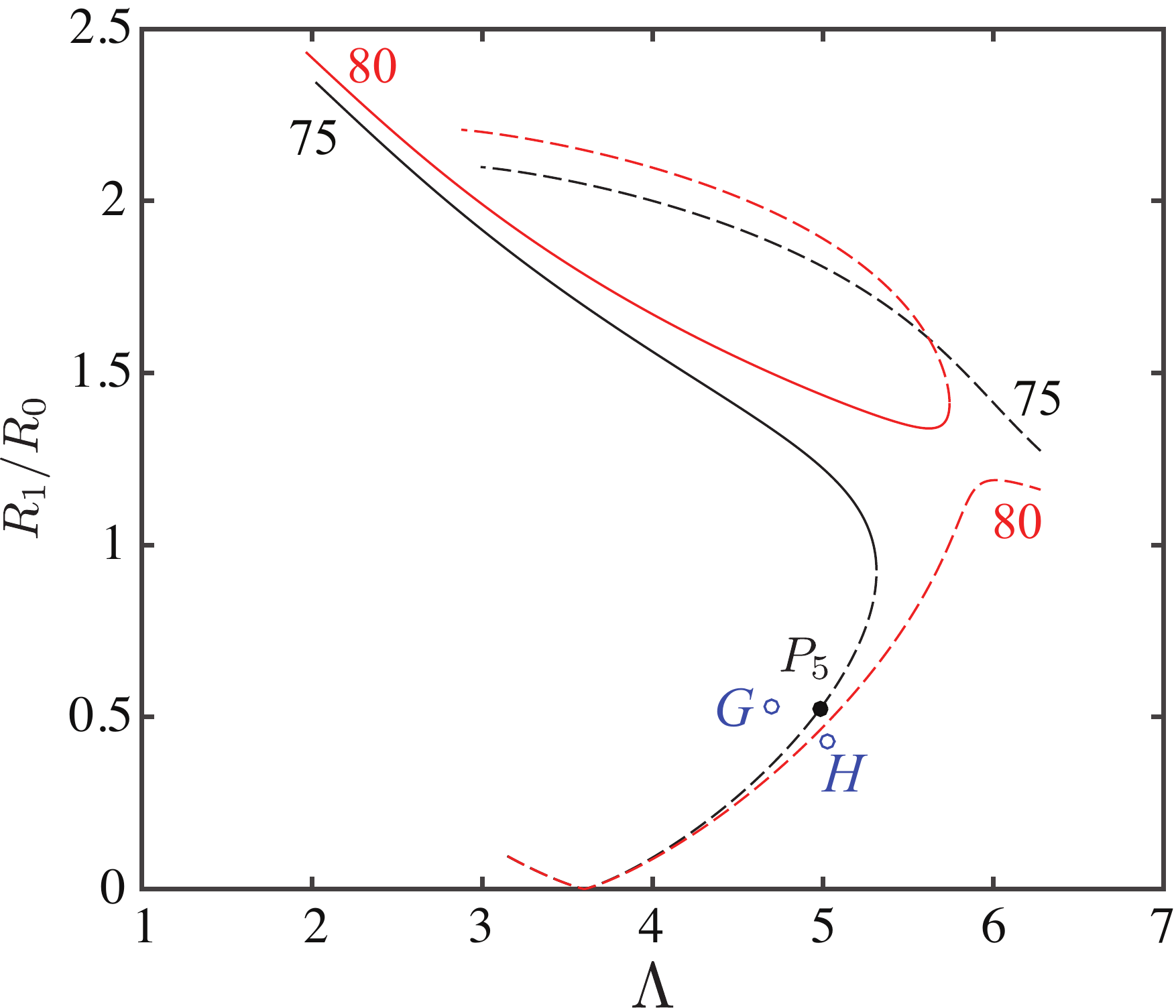}
 \caption{Free-contact line equilibrium branches of a fixed-volume ($v^{*}=7.171$) liquid bridge in the vicinity of the transcritical bifurcation, indicating stable (solid) and unstable (dashed) states. Numeric labels denote the contact angle $\theta_{c}$ in degrees. Circles indicate the states at $G$, $H$, and $P_5$ as the bridge approaches breakup in Fig.~\ref{fig4:figure10}.}
\label{fig4:figure13}
 \end{figure}
 
We further elaborate on the behaviour of the contact-line radius after stability loss by constructing the equilibrium branches at contact angles near $\theta_{r2}$. The scaled form of the slender-jet approximation~\citep{eggers1994drop} suggests that, except very close to the singularity where the capillary number scales as $\mbox{Ca}\sim O(1)$, the bridge profile can be reasonably approximated by the Young-Laplace equation. This approximation is expected---also experimentally shown by \cite{qian2011motion}---to be accurate for volume-controlled stretching since the liquid velocity inside the bridge is restricted by the volume constraint. Therefore, the bridge must evolve along unstable equilibrium branches at fixed volume. We apply this approximation as a guide to investigate the bridge evolution after stability loss and away from the pinch-off (\eg, bridge $H$ in Fig.~\ref{fig4:figure10}) and provide a better understanding of the relationship between the receding contact angle and dispensed-drop volume. 

 \citet{akbari2014bridge} showed that, at fixed $\theta_{c}$, liquid bridges with a free contact line exhibit a transcritical bifurcation at a point along the lower boundary of the stability region. Transcritical bifurcations were represented by equilibrium branches at fixed $\theta_{c}$ and $\Lambda$ in pressure versus volume diagrams. Figure~\ref{fig4:figure13} shows equilibrium branches in the vicinity of a transcritical bifurcation in $R_1/R_0$ versus $\Lambda$ diagram at fixed $v^{*}$ and $\theta_{c}$. As discussed for Figs.~\ref{fig4:figure12}b and \ref{fig4:figure11}c, the bridge after the pinned-pinned to pinned free transition at $P_5$ lies on an unstable segment of the primary branch at $\theta_{c}=75^{\circ}$. Upon stability loss, the dynamic contact angle increases from $\approx 75$ to $81^{\circ}$ with the bridge moving from the primary branch of $\theta_{c}=75^{\circ}$ at $P_5$ to the secondary branch of $\theta_{c}=81^{\circ}$ at $H$. Here, the opposite behaviour of $R_1$ with $\Lambda$ along stable and unstable branches is notable: the contact-line radius decreases (increases) during stretching along stable (unstable) branches, favouring small (large) drops. This implies that  evolution upon the loss of stability at a fixed-contact-angle leads to larger drops; thus, to achieve smaller drops, the dynamic contact angle must increase following stability loss, upon which the contact line accelerates toward pinch-off. Therefore, there are cases where \emph{a dynamic contact angle that deviates from the receding contact angle plays a key role in determining whether the contact line expands or contracts.}
 
 \section{Concluding remarks} \label{sec:conclusion}
 
We have experimentally and theoretically studied the stability and breakup of weightless liquid bridges on surfaces with contact-angle hysteresis. Experiments were performed in a Plateau tank where the effect of gravity was alleviated by density matching. To achieve free and pinned contact lines, the contact angle was adjusted by adding SDS to the bath. For liquid bridges with a free contact line, experimental measurements validated the theoretical predictions of the stability limits. At fixed volume, liquid bridges with a free contact line exhibit a larger breakup length than those with a pinned contact line, demonstrating the destabilizing effect of a free contact line, as theoretically predicted by \citet{akbari2014bridge}. 
 
 We examined the effect of contact-angle hysteresis on the maximum-length stability limit, showing that the breakup length can not always be associated with the static stability limit. Depending on the drop volume and receding contact angle, liquid bridges may lose stability during the pinned-pinned to pinned-free transition at an unstable state away from its critical state. This has significant implications for the dynamics following stability loss. Unstable states that are far from critical states generally have large instability margins, which can transform potential energy to kinetic energy upon stability loss, having a significant impact on the breakup dynamics and dispensed-drop size. Constructing equilibrium branches in the vicinity of the receding contact angle revealed that a complex interplay between the dynamic contact angle (determined by its speed), receding contact angle when losing stability, and bridge volume determine whether the contact line is expanding or contracting upon stability loss.  Furthermore, we showed (by one example) that the contact-line motion upon stability loss at fixed contact angle can be advancing, thus disfavouring small-drop deposition. 
 
 More comprehensive computational investigations of the stability and breakup of liquid bridges will hopefully provide deeper insights into the relationship between the contact-angle hysteresis and dispensed-drop size, and on the effectiveness of surface hydrophobization (to modulate the contact angle) for micro-deposition. 
 
 \section*{Acknowledgements} \label{sec:acknowledgements}

Supported by the NSERC Innovative Green Wood Fibre Products Network (R.J.H.), and a McGill Engineering Doctoral Award (A.A.). The authors acknowledge Jessie Zhang for assistance performing the experiments and image analysis. 

\bibliography{mybib}

\begin{thebibliography}{46}
\expandafter\ifx\csname natexlab\endcsname\relax\def\natexlab#1{#1}\fi

\bibitem[Akbari {\em et~al.\/}(2015{\natexlab{{\em a\/}}})Akbari, Hill \&
  van~de Ven]{akbari2014catenoid}
{\sc Akbari, A., Hill, R.~J. \& van~de Ven, T. G.~M.} 2015{\natexlab{{\em
  a\/}}} Catenoid stability with a free contact line. {\em Accepted in SIAM J.
  Appl. Math.\/} .

\bibitem[Akbari {\em et~al.\/}(2015{\natexlab{{\em b\/}}})Akbari, Hill \&
  van~de Ven]{akbari2014an}
{\sc Akbari, A., Hill, R.~J. \& van~de Ven, T. G.~M.} 2015{\natexlab{{\em
  b\/}}} An elastocapillary model of wood-fibre collapse. {\em Proc. R. Soc.
  A\/} {\bf 471}~(2179).

\bibitem[Akbari {\em et~al.\/}(2015{\natexlab{{\em c\/}}})Akbari, Hill \&
  van~de Ven]{akbari2014bridge}
{\sc Akbari, A., Hill, R.~J. \& van~de Ven, T. G.~M.} 2015{\natexlab{{\em
  c\/}}} Liquid bridge breakup in contact-drop dispensing: Liquid bridge
  stability with a free contact line. {\em Accepted in Phys. Rev. E\/} .

\bibitem[Akbari {\em et~al.\/}(2015{\natexlab{{\em d\/}}})Akbari, Hill \&
  van~de Ven]{akbari2014stability}
{\sc Akbari, A., Hill, R.~J. \& van~de Ven, T. G.~M.} 2015{\natexlab{{\em
  d\/}}} Stability and folds in an elastocapillary system. {\em Submitted to
  SIAM J. Appl. Math.\/} .

\bibitem[Bachmann {\em et~al.\/}(2000)Bachmann, Ellies \&
  Hartge]{bachmann2000development}
{\sc Bachmann, J., Ellies, A. \& Hartge, K.~H.} 2000 Development and
  application of a new sessile drop contact angle method to assess soil water
  repellency. {\em J. Hydrology\/} {\bf 231}, 66--75.

\bibitem[Bezdenejnykh {\em et~al.\/}(1992)Bezdenejnykh, Meseguer \&
  Perales]{bezdenejnykh1992experimental}
{\sc Bezdenejnykh, N.~A., Meseguer, J. \& Perales, J.~M.} 1992 Experimental
  analysis of stability limits of capillary liquid bridges. {\em Phys. Fluids
  A\/} {\bf 4}~(4), 677--680.

\bibitem[Bezdenejnykh {\em et~al.\/}(1999)Bezdenejnykh, Meseguer \&
  Perales]{bezdenejnykh1999experimental}
{\sc Bezdenejnykh, N.~A., Meseguer, J. \& Perales, J.~M.} 1999 An experimental
  analysis of the instability of nonaxisymmetric liquid bridges in a
  gravitational field. {\em Phys. Fluids A\/} {\bf 11}~(10), 3181--3185.

\bibitem[Boudaoud {\em et~al.\/}(2007)Boudaoud, Bico \&
  Roman]{boudaoud2007elastocapillary}
{\sc Boudaoud, A., Bico, J. \& Roman, B.} 2007 Elastocapillary coalescence:
  {A}ggregation and fragmentation with a maximal size. {\em Phys. Rev. E.\/}
  {\bf 76}~(6), 060102.

\bibitem[Chandra \& Yang(2009)]{chandra2009capillary}
{\sc Chandra, D. \& Yang, S.} 2009 Capillary-force-induced clustering of
  micropillar arrays: is it caused by isolated capillary bridges or by the
  lateral capillary meniscus interaction force? {\em Langmuir\/} {\bf 25}~(18),
  10430--10434.

\bibitem[Chen {\em et~al.\/}(2013)Chen, Amirfazli \& Tang]{chen2013modeling}
{\sc Chen, H., Amirfazli, A. \& Tang, T.} 2013 Modeling liquid bridge between
  surfaces with contact angle hysteresis. {\em Langmuir\/} {\bf 29}~(10),
  3310--3319.

\bibitem[Chen {\em et~al.\/}(2014)Chen, Tang \& Amirfazli]{chen2014liquid}
{\sc Chen, H., Tang, T. \& Amirfazli, A.} 2014 Liquid transfer mechanism
  between two surfaces and the role of contact angles. {\em Soft Matter\/} {\bf
  10}~(15), 2503--2507.

\bibitem[Cohen \& Mahadevan(2003)]{cohen2003kinks}
{\sc Cohen, A.~E. \& Mahadevan, L.} 2003 Kinks, rings, and rackets in
  filamentous structures. {\em P. Natl. Acad. Sci. USA\/} {\bf 100}~(21),
  12141--12146.

\bibitem[Dodds {\em et~al.\/}(2009)Dodds, da~Silveira~Carvalho \&
  Kumar]{dodds2009stretching}
{\sc Dodds, S., da~Silveira~Carvalho, M. \& Kumar, S.} 2009 Stretching and
  slipping of liquid bridges near plates and cavities. {\em Phys. Fluids\/}
  {\bf 21}, 092103.

\bibitem[Drelich {\em et~al.\/}(1996)Drelich, Wilbur, Miller \&
  Whitesides]{drelich1996contact}
{\sc Drelich, J., Wilbur, J.~L., Miller, J.~D. \& Whitesides, G.~M.} 1996
  Contact angles for liquid drops at a model heterogeneous surface consisting
  of alternating and parallel hydrophobic/hydrophilic strips. {\em Langmuir\/}
  {\bf 12}~(7), 1913--1922.

\bibitem[Eggers(1997)]{eggers1997nonlinear}
{\sc Eggers, J.} 1997 Nonlinear dynamics and breakup of free-surface flows.
  {\em Rev. Mod. Phys.\/} {\bf 69}~(3), 865.

\bibitem[Eggers \& Dupont(1994)]{eggers1994drop}
{\sc Eggers, J. \& Dupont, T.~F.} 1994 Drop formation in a one-dimensional
  approximation of the {N}avier--{S}tokes equation. {\em J. Fluid Mech.\/} {\bf
  262}, 205--221.

\bibitem[Farshid-Chini \& Amirfazli(2010)]{chini2010understanding}
{\sc Farshid-Chini, S. \& Amirfazli, A.} 2010 Understanding pattern collapse in
  photolithography process due to capillary forces. {\em Langmuir\/} {\bf
  26}~(16), 13707--13714.

\bibitem[Gao \& McCarthy(2006)]{gao2006contact}
{\sc Gao, L. \& McCarthy, T.~J.} 2006 Contact angle hysteresis explained. {\em
  Langmuir\/} {\bf 22}~(14), 6234--6237.

\bibitem[Garc{\'\i}a~Velarde(1988)]{velarde1988physicochemical}
{\sc Garc{\'\i}a~Velarde, M.} 1988 {\em Physicochemical Hydrodynamics:
  Interfacial Phenomena\/}, , vol. 174. Springer.

\bibitem[Huo {\em et~al.\/}(2008)Huo, Zheng, Zheng, Giam, Zhang \&
  Mirkin]{huo2008polymer}
{\sc Huo, F., Zheng, Z., Zheng, G., Giam, L.~R., Zhang, H. \& Mirkin, C.~A.}
  2008 Polymer pen lithography. {\em Science\/} {\bf 321}~(5896), 1658--1660.

\bibitem[Joanny \& de~Gennes(1984)]{joanny1984model}
{\sc Joanny, J.~F. \& de~Gennes, P.-G.} 1984 A model for contact angle
  hysteresis. {\em J Chem. Phys.\/} {\bf 81}~(1), 552--562.

\bibitem[Kim \& Mahadevan(2006)]{kim2006capillary}
{\sc Kim, H. \& Mahadevan, L.} 2006 Capillary rise between elastic sheets. {\em
  J. Fluid Mech.\/} {\bf 548}, 141--150.

\bibitem[Kwon {\em et~al.\/}(2008)Kwon, Kim, Pu{\"e}ll \&
  Mahadevan]{kwon2008equilibrium}
{\sc Kwon, H., Kim, H., Pu{\"e}ll, J. \& Mahadevan, L.} 2008 Equilibrium of an
  elastically confined liquid drop. {\em J. Appl. Phys.\/} {\bf 103}~(9),
  093519.

\bibitem[Lowry \& Steen(1994)]{lowry1994stabilization}
{\sc Lowry, B.~J. \& Steen, P.~H.} 1994 Stabilization of an axisymmetric liquid
  bridge by viscous flow. {\em Int. J. Multiphas. Flow\/} {\bf 20}~(2),
  439--443.

\bibitem[Lucassen-Reynders {\em et~al.\/}(1981)Lucassen-Reynders, Lucassen \&
  Giles]{lucassen1981surface}
{\sc Lucassen-Reynders, E.~H., Lucassen, J. \& Giles, D.} 1981 Surface and bulk
  properties of mixed anionic/cationic surfactant systems i. equilibrium
  surface tensions. {\em J. Colloid Interf. Sci.\/} {\bf 81}~(1), 150--157.

\bibitem[Lutfurakhmanov {\em et~al.\/}(2010)Lutfurakhmanov, Loken, Schulz \&
  Akhatov]{lutfurakhmanov2010capillary}
{\sc Lutfurakhmanov, A., Loken, G.~K., Schulz, D.~L. \& Akhatov, I.~S.} 2010
  Capillary-based liquid microdroplet deposition. {\em Appl. Phys. Lett.\/}
  {\bf 97}~(12), 124107.

\bibitem[Marr \& Hildreth(1980)]{marr1980theory}
{\sc Marr, D. \& Hildreth, E.} 1980 Theory of edge detection. {\em Proc. R.
  Soc. B\/} {\bf 207}~(1167), 187--217.

\bibitem[Mart{\'\i}nez \& Perales(1986)]{martinez1986liquid}
{\sc Mart{\'\i}nez, I. \& Perales, J.~M.} 1986 Liquid bridge stability data.
  {\em J. Cryst. Growth\/} {\bf 78}~(2), 369--378.

\bibitem[Mastrangelo \& Hsu(1993)]{mastrangelo1993mechanical}
{\sc Mastrangelo, C.~H. \& Hsu, C.~H.} 1993 Mechanical stability and adhesion
  of microstructures under capillary forces. {I}. {B}asic theory. {\em J.
  Microelectromech. S.\/} {\bf 2}~(1), 33--43.

\bibitem[Meseguer {\em et~al.\/}(1995)Meseguer, Slobozhanin \&
  Perales]{meseguer1995review}
{\sc Meseguer, J., Slobozhanin, L.~A. \& Perales, J.~M.} 1995 A review on the
  stability of liquid bridges. {\em Adv. Space Res.\/} {\bf 16}~(7), 5--14.

\bibitem[Mukerjee \& Mysels(1971)]{mukerjee1971critical}
{\sc Mukerjee, P. \& Mysels, K.~J.} 1971 Critical micelle concentrations of
  aqueous surfactant systems. {\em Tech. Rep.\/}. DTIC Document.

\bibitem[Myshkis(1965)]{Myshkis1965on}
{\sc Myshkis, A.} 1965 On depressions. {\em {USSR} Comput. Math. Math. Phys.\/}
  {\bf 5}~(3), 193 -- 201.

\bibitem[Myshkis {\em et~al.\/}(1987)Myshkis, Babskii, Kopachevskii,
  Slobozhanin, Tyuptsov \& Wadhwa]{myshkis1987low}
{\sc Myshkis, A.~D., Babskii, V.~G., Kopachevskii, N.~D., Slobozhanin, L.~A.,
  Tyuptsov, A.~D. \& Wadhwa, R.~S.} 1987 {\em Low-gravity fluid mechanics\/}.
  Springer-Verlag Berlin.

\bibitem[Piner {\em et~al.\/}(1999)Piner, Zhu, Xu, Hong \&
  Mirkin]{piner1999dip}
{\sc Piner, R.~D., Zhu, J., Xu, F., Hong, S. \& Mirkin, C.~A.} 1999 Dip-pen
  nanolithography. {\em Science\/} {\bf 283}~(5402), 661--663.

\bibitem[Plateau(1873)]{plateau1873statique}
{\sc Plateau, J. A.~F.} 1873 {\em Statique exp{\'e}rimentale et th{\'e}orique
  des liquides soumis aux seules forces mol{\'e}culaires\/}. Gauthier-Villars.

\bibitem[Pokroy {\em et~al.\/}(2009)Pokroy, Kang, Mahadevan \&
  Aizenberg]{pokroy2009self}
{\sc Pokroy, B., Kang, S.~H., Mahadevan, L. \& Aizenberg, J.} 2009
  Self-organization of a mesoscale bristle into ordered, hierarchical helical
  assemblies. {\em Science\/} {\bf 323}~(5911), 237--240.

\bibitem[Qian \& Breuer(2011)]{qian2011motion}
{\sc Qian, B. \& Breuer, K.~S.} 2011 The motion, stability and breakup of a
  stretching liquid bridge with a receding contact line. {\em J. Fluid Mech.\/}
  {\bf 666}, 554--572.

\bibitem[Qian {\em et~al.\/}(2009)Qian, Loureiro, Gagnon, Tripathi \&
  Breuer]{qian2009micron}
{\sc Qian, B., Loureiro, M., Gagnon, D.~A., Tripathi, A. \& Breuer, K.~S.} 2009
  Micron-scale droplet deposition on a hydrophobic surface using a retreating
  syringe. {\em Phys. Rev. Lett.\/} {\bf 102}~(16), 164502.

\bibitem[Rayleigh(1879{\natexlab{{\em a\/}}})]{rayleigh1879capillary}
{\sc Rayleigh, L.} 1879{\natexlab{{\em a\/}}} On the capillary phenomena of
  jets. {\em Proc. R. Soc. A\/} {\bf 29}~(196-199), 71--97.

\bibitem[Rayleigh(1879{\natexlab{{\em b\/}}})]{rayleigh1879instability}
{\sc Rayleigh, L.} 1879{\natexlab{{\em b\/}}} On the instability of jets. {\em
  Proc. R. Soc. A\/} {\bf 10}, 4--13.

\bibitem[Russo \& Steen(1986)]{Russo1986154}
{\sc Russo, M.~J. \& Steen, P.~H.} 1986 Instability of rotund capillary bridges
  to general disturbances: Experiment and theory. {\em J. Colloid Interf.
  Sci.\/} {\bf 113}~(1), 154 -- 163.

\bibitem[Salaita {\em et~al.\/}(2007)Salaita, Wang \&
  Mirkin]{salaita2007applications}
{\sc Salaita, K., Wang, Y. \& Mirkin, C.~A.} 2007 Applications of dip-pen
  nanolithography. {\em Nat. Nanotechnol.\/} {\bf 2}~(3), 145--155.

\bibitem[Sanz \& Martinez(1983)]{sanz1983minimum}
{\sc Sanz, A. \& Martinez, I.} 1983 Minimum volume for a liquid bridge between
  equal disks. {\em J. Colloid Interf. Sci.\/} {\bf 93}~(1), 235--240.

\bibitem[Shim {\em et~al.\/}(2011)Shim, Braunschweig, Liao, Chai, Lim, Zheng \&
  Mirkin]{shim2011hard}
{\sc Shim, W., Braunschweig, A.~B., Liao, X., Chai, J., Lim, J.~K., Zheng, G.
  \& Mirkin, C.~A.} 2011 Hard-tip, soft-spring lithography. {\em Nature\/} {\bf
  469}~(7331), 516--520.

\bibitem[Slobozhanin {\em et~al.\/}(2002)Slobozhanin, Alexander \&
  Patel]{slobozhanin2002stability}
{\sc Slobozhanin, L.~A., Alexander, J. I.~D. \& Patel, V.~D.} 2002 The
  stability margin for stable weightless liquid bridges. {\em Phys. Fluids\/}
  {\bf 14}, 209--224.

\bibitem[Slobozhanin {\em et~al.\/}(1997)Slobozhanin, Alexander \&
  Resnick]{slobozhanin1997bifurcation}
{\sc Slobozhanin, L.~A., Alexander, J. I.~D. \& Resnick, A.~H.} 1997
  Bifurcation of the equilibrium states of a weightless liquid bridge. {\em
  Phys. Fluids\/} {\bf 9}, 1893--1905.

\end{thebibliography}
\bibliographystyle{jfm}

\end{document}